\documentclass[pdflatex,sn-vancouver,iicol]{sn-jnl}

\usepackage{siunitx}
\jyear{2021}%

\theoremstyle{thmstyleone}%
\theoremstyle{thmstyletwo}%

\theoremstyle{thmstylethree}%

\usepackage{siunitx}
\usepackage[switch]{lineno}
\newcommand*\patchAmsMathEnvironmentForLineno[1]{
  \expandafter\let\csname old#1\expandafter\endcsname\csname #1\endcsname
  \expandafter\let\csname oldend#1\expandafter\endcsname\csname end#1\endcsname
  \renewenvironment{#1}
  {\linenomath\csname old#1\endcsname}
  {\csname oldend#1\endcsname\endlinenomath}}
  \newcommand*\patchBothAmsMathEnvironmentsForLineno[1]{
  \patchAmsMathEnvironmentForLineno{#1}
  \patchAmsMathEnvironmentForLineno{#1*}}
  \AtBeginDocument{
  \patchBothAmsMathEnvironmentsForLineno{equation}
  \patchBothAmsMathEnvironmentsForLineno{align}
  \patchBothAmsMathEnvironmentsForLineno{flalign}
  \patchBothAmsMathEnvironmentsForLineno{alignat}
  \patchBothAmsMathEnvironmentsForLineno{gather}
  \patchBothAmsMathEnvironmentsForLineno{multline}
}

\begin{document}

\title{Squeezed Ground States in a Spin-1 Bose-Einstein Condensate}


\author{\fnm{Lin} \sur{Xin}}

\author{\fnm{Maryrose} \sur{Barrios}}

\author{\fnm{Julia T.} \sur{Cohen}}

\author{\fnm{Michael S. } \sur{Chapman}}

\affil{\orgdiv{School of Physics}, \orgname{Georgia Institute of Technology}, \orgaddress{\city{Atlanta}, \postcode{30332}, \state{GA}, \country{U.S.A}}}


\abstract{We generate spin squeezed ground states in an atomic spin-1  Bose-Einstein condensate tuned near the quantum critical point between the polar and ferromagnetic quantum phases of the interacting spin ensemble. In contrast to typical non-equilibrium methods for preparing atomic squeezed states by quenching through a quantum phase transition, squeezed ground states are time-stationary and remain squeezed for the lifetime of the condensate. A squeezed ground state with a metrological improvement up to 6-8 dB and a constant squeezing angle maintained over 2~s is demonstrated.}

\maketitle

For quantum-limited metrology with $N$ uncorrelated particles in an atomic clock or optical interferometer for example,  the uncertainty principle provides the standard quantum limit (SQL) of relative measurement precision, $1/\sqrt{N}$. An important frontier of research in metrology is the development of techniques to surpass this limit using quantum squeezed states or other entangled states  \cite{Smerzi2018,MA201189}. These techniques are expected to play an important role in the next generation of quantum sensors  \cite{Pedrozo-Penafiel2020,PhysRevLett.125.100402,Aasi2013}. 
Atomic Bose-Einstein condensates (BECs) with internal spin degrees of freedom are a promising platform for creating and characterizing atomic spin squeezed and other entangled states \cite{Stamper-Kurn,Kawaguchi2012}. These systems feature strong collisional spin interactions, tunable Hamiltonians with quantum phase transitions (QPT) and low-noise tomographic quantum spin state measurement capabilities that allow exploration of a wide range of interesting phenomena including squeezing \cite{Esteve2008,Hamley2012,PhysRevLett.112.155304}, dynamical stabilization \cite{PhysRevLett.111.090403}, parametric excitation \cite{Hoang2016}, and studies of the quantum phase transition \cite{Hoang2017,PhysRevLett.107.195306,Luo620} including Kibble-Zurek universality \cite{Anquez2016}. 
Experimental demonstrations of collisionally-induced spin squeezing in condensates have mainly utilized non-equilibrium many-body dynamical  evolution in one-axis twisting or similar Hamiltonians  \cite{Kitagawa1993,Gross2010, Hamley2012,Muessel2015} following a deep quench across the QPT from an initially uncorrelated state; recently spin squeezed states have also been generated using parametric/Floquet excitation \cite{Hoang2016,PhysRevLett.125.033401}.

In contrast to entanglement and squeezing in excited states, there is much interest in studying similar phenomenon in the ground states. Entangled ground states are central to adiabatic quantum computing and understanding strong-correlated many-body systems, and there are also compelling applications to quantum enhanced metrology \cite{doi:10.1080/0950034021000011536}. To this last point, there have been experiments using adiabatic \cite{Hoang2017} or quasi-adiabatic \cite{Luo620,Zou6381} evolution across the symmetry-breaking phase transition to create highly-entangled states such as Dicke states and twin-Fock states \cite{Zhang2013}.

The focus of this paper is the creation and investigation of Gaussian squeezed ground states. These states arise naturally as the Hamiltonian is tuned near the symmetry-breaking QPT and offer the advantage that the squeezed state properties are determined by the properties of the final Hamiltonian rather than the details of the non-equilibrium evolution and are thus easier to characterize and control. In particular, the minimum squeezed quadrature angle for the ground state has a fixed orientation independent of the Hamiltonian parameters such as density and magnetic field. In contrast, the minimum squeezing quadrature angle in non-equilibrium methods is both time and atom number dependent \cite{Hamley2012}, which poses serious challenges for highly squeezed states. Finally, spin squeezed ground states provide opportunities to more carefully investigate long-term evolution of entanglement in spin ensembles because the squeezing is now in a stationary state.
A distinguishing feature of the investigation described in this letter is the use of a double-quench shortcut \cite{xin2021fast} to approach the QPT that significantly shortens the state preparation time compared to adiabatic methods.  Decreasing the preparation time improves both the fidelity of the target state and the detection limit due to uncorrelated atom losses. 

\begin{figure*}
\includegraphics[width=1\textwidth]{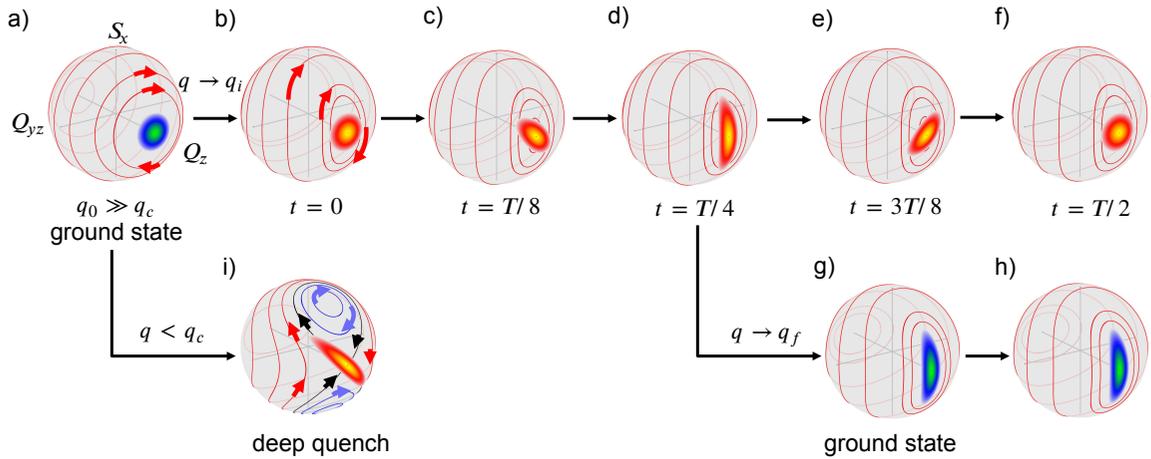}
\caption{\label{fig1} \textbf{The spin-1 states in the $\hat{S}_z=0$ subspace and their evolution can be visualized on a $\{S_{x}, Q_{yz}, Q_z\}$ Bloch sphere.} \textbf{(a)} The initial state is an uncorrelated ground state at $q\gg q_c$ with symmetric uncertainties in $S_x$ and $Q_{yz}$. \textbf{(b)-(f)} following a sudden quench to $q_i\gtrsim q_c$ at $t=0$, the ground state remains polar, but the fluctuations evolve periodically along elliptical orbits with a frequency $\omega_i= 2\pi/T$. \textbf{(g)-(h)} A second quench at $T/4$ to a suitably chosen $q_f$ will de-excite the condensate into a stationary squeezed ground state.  \textbf{(i)} Standard non-equilibrium method of generating spin-1 squeezing following a sudden deep quench across the QCP to the FM phase \cite{Hamley2012,doi:10.1126/science.1250147}. } 
\end{figure*}

The spin dynamics of a small spin-1 condensate in a magnetic field oriented along the $z$ direction  are described by the Hamiltonian \cite{Hamley2012}:
 \begin{equation}
\hat{H}=\frac{c}{2N}\hat{S}^2-\frac{q}{2}\hat{Q}_{z},
\label{hamiltonianreduced}
\end{equation}
where $\hat{S}$ is the collective spin operator, and $\hat{Q}_{z}$ is a collective nematic/quadrupole operator. The coefficient $c/2N$ is the collisional spin interaction energy per particle,  and $q \propto B^2$ is the quadratic Zeeman energy per particle. For the $^{87}\mbox{Rb}$ $F=1$ hyperfine state, $c<0$ meaning the condensate has a ferromagnetic (FM) phase and a polar phase, separated by a QCP at $q=2\vert c\vert  \equiv q_c$ (see Supplementary Section I).    

\begin{figure*}
\includegraphics[width=1\textwidth]{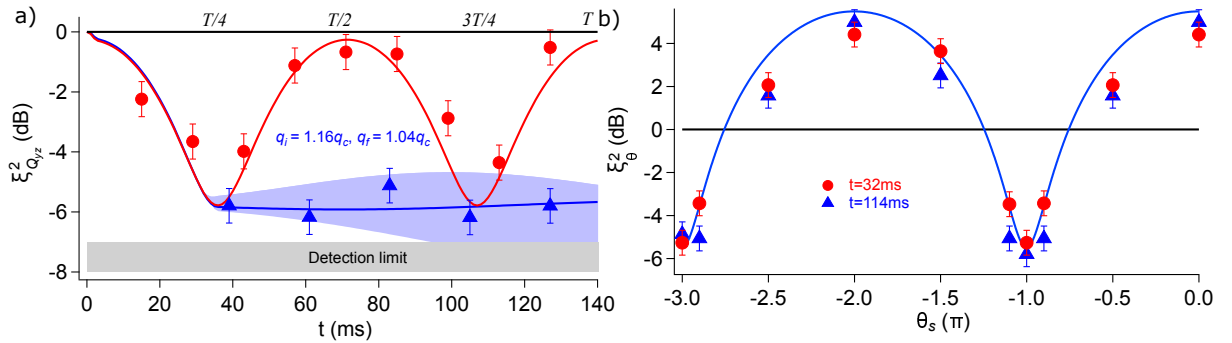}
\caption{\label{fig2}\textbf{Time-stationary squeezing and periodic squeezing.} \textbf{(a)} Measurement of time-stationary squeezing in the $\Delta Q_{yz}$ observable following the double quench sequence $q_0 \to q_i \to q_f$ designed to create a squeezed ground state at $q_f$ (blue triangles). These data are compared to a single quench $q_0 \to q_i$ (red circles), which exhibit periodic squeezing and unsqueezing in $\Delta Q_{yz}$.  Simulation results with $c=-8.2\pm 0.1$ Hz (blue shaded area) are compared with the data.  \textbf{(b)} Tomographic measurements of the fluctuations at $t=T/4$ (red circles) and at a much later time ($t\sim 3T/4$) after the second quench (blue triangles).  The error bars indicate the standard deviation of measured variance determined from 100 repeated measurements per data point.}
\end{figure*}

We begin by describing the basic idea behind the experiment. The starting point is a spin-1 condensate prepared in the  $m_F =0$ Zeeman state  at a high magnetic field such that $q=q_0\gg q_c$ and  the spin interaction term of the Hamiltonian can be ignored. This is an uncorrelated ground state with Heisenberg uncertainty for the complimentary observables $\Delta S_{x}\Delta Q_{yz}=N$, where $\hat{S_x}$ is the collective spin operator in $x$ direction, and $\hat{Q}_{yz}$ is the collective nematic operator between $y$ and $z$ direction. Throughout the text, operators are indicated by carets, while the corresponding symbol without the caret indicated their expectation value.  The phase space of the system can be visualized on a Bloch sphere of $\{S_{x},Q_{yz},Q_z\}$  (see Fig. \ref{fig1}) where the ground state is located at the $Q_z=1$ pole with symmetric uncertainties in $S_x$ and $Q_{yz}$. In earlier demonstrations of spin-nematic squeezing \cite{Hamley2012,doi:10.1126/science.1250147}, the squeezing was generated by non-equilibrium evolution from an unstable fixed point following a deep quench across the QCP to the FM phase as shown in Fig. \ref{fig1}(i). In this work, we are interested in creating squeezing in the polar phase in the neighborhood of the QCP and, in particular, creating squeezing in the ground state of the system with $q \gtrsim  q_c$. We again begin with a sudden quench from $q_0$, but now to a final field above the QCP,  $q_i  \gtrsim q_c$. At this field,  the ground state remains polar in character, but the spin interactions are no longer negligible and  distort the semi-classical orbits of the system into ellipses. Subsequent evolution of the initially symmetric uncertainties gives rise to periodic squeezing and unsqueezing with a frequency $\omega_i=\sqrt{q_i(q_i-q_c)}$ as shown in Fig. \ref{fig1}(b)-(f) from the energy gap \cite{Hoang2016}. Of course, this is an excited state of the system with dynamically evolving observables, in this case the uncertainties $\Delta S_x$ and $ \Delta Q_{yz}$.  Although this state is not a ground state of the Hamiltonian $\hat{H}(q_i)$, it is the ground state of another Hamiltonian $\hat{H}(q_f)$ where $q_i> q_f > q_c$. To end with the condensate in a ground state, we perform a second quench with a timing and final field value chosen to match the evolving state with the shape of the ground state of the final Hamiltonian. This second quench results in the system in the ground state of $\hat{H}(q_f)$  as shown in  Fig. \ref{fig1}(g)-(h).

The ground state of $\hat{H}(q_f)$ exhibits squeezing in the variance of $Q_{yz}$  by an amount \cite{xin2021fast}:
\begin{equation}
\xi^{2}_{Q_{yz}}= \Delta Q_{yz}^2/N = 1/\eta,
\label{squeezing_amount}
\end{equation}
where $1/ \eta=\sqrt{1-q_c/q_f}$, and anti-squeezing by an amount $\eta$ in the complimentary observable $S_x$. In order to end in the ground state, the second quench needs to occur at a time $T/4=\pi/(2\omega_i)$ and $q_f$ needs to satisfy the relation $(q_i-q_c)/q_i=1/\eta$.  Of course, it is also possible to adiabatically ramp the Hamiltonian directly from $q_0 \rightarrow q_f$, but the double quench shortcut method is at least $\sqrt{\eta}$ faster than the shortest adiabatic ramp time $T_{adiab}\geq 2\pi\eta/{q_f}$ (see \cite{xin2021fast} for details).

We now turn to the experimental measurements.  We first investigate the single quench non-equilibrium periodic squeezing following Fig. \ref{fig1}(b)-(f). A condensate of 50k atoms is prepared in the $m_F =0$ state in an optical dipole cross trap at a high field, $q_0= 5q_c $. Following a sudden quench to $q_i = 1.16q_c$, the condensate is allowed to freely evolve. The mean spin populations do not significantly change as the condensate is still in the polar phase, however the spin fluctuations do evolve. In Fig. \ref{fig2}(a), measurements of the time evolution of $\Delta Q_{yz}$ are shown that exhibit periodic squeezing and unsqueezing; measurements of $\Delta S_{x}$ show complimentary behavior of periodic anti-squeezing (see Supplementary Section II). In Fig. \ref{fig2}(b), tomographic measurements of the fluctuations at the point of maximum $ Q_{yz}$ squeezing ($t=T/4$) are shown. Each data point corresponds to a measurement at a different quadrature phase $\theta = \theta_s/2$, where $\theta_s$ is the relative phase between $m_F=0$ and $m_F=\pm1$ spin components:
\begin{equation}
    \xi^2_\theta=\Delta(S_x\cos \theta+Q_{yz}\sin \theta)^2/N.
\end{equation} 
The data show up to $-6$ dB of squeezing and symmetric anti-squeezing. The data are compared with simulations that show good qualitative agreement; however, it is necessary to scale the simulations by    $\xi^2 = (\xi^2_{sim})^{0.7}$ to quantitatively match the observed squeezing --- possible explanations are discussed in the Supplementary Section I. In the figures throughout, the simulations are scaled  to account for this discrepancy.

Also shown in Fig. \ref{fig2} are  data taken following the double quench sequence $q_0 \to q_i \to q_f$ designed to achieve the squeezed ground state of $\hat{H}(q_f)$.  In Fig. \ref{fig2}(a), the data show that following the second quench to $q_f=1.04q_c$, the time evolution of $\Delta Q_{yz}$ remains constant at the level of the maximum squeezing previously observed, as expected for the ground state. The data are compared with a simulation result including a $\pm 0.1$ Hz uncertainty in $c$ (see Methods). The precise values of $T$ and $q_f$ are determined from the single quench data.
Tomographic measurements of the fluctuations of the ground state shown in Fig. \ref{fig2}(b) taken at a much later time ($t \sim 3T/4$), are indistinguishable from measurements made of the periodic squeezing at ($t = T/4$), as expected. 
Furthermore, in addition to a constant squeezing amplitude, the maximum squeezing angle (the minimum quadrature angle) $\theta_{s,min}=\min\{\xi^2_\theta\vert\theta_s\}=-\pi$ remains constant following the second quench. This is in stark contrast to the deep quench method (Fig.~\ref{fig1}(i)) for which  $\theta_{s,min}$ is a function of $c,q$ and evolves dynamically (see Supplementary Section I). The experimental data is corrected for the photon shot noise and the background imaging noise and the   detection limit of the squeezing is $-7$ dB (Methods). From the measurement of $-6$ dB of squeezing, it is possible to determine the entanglement breadth of the spin ensemble \cite{PhysRevLett.86.4431,PhysRevLett.112.155304,Zou6381}. From this, we can conclude that a non-separable (entangled) subset of 600 particles is detected in the squeezed ground state (Supplementary Section II). For comparison, we have also used an adiabatic ramp method to create the squeezed ground state (see Supplementary Section II). It is clear that the double quench method is superior, offering  $\geq \sqrt{\eta}$ faster preparation and higher squeezing by minimizing atomic losses.

\begin{figure}
\includegraphics[width=0.5\textwidth]{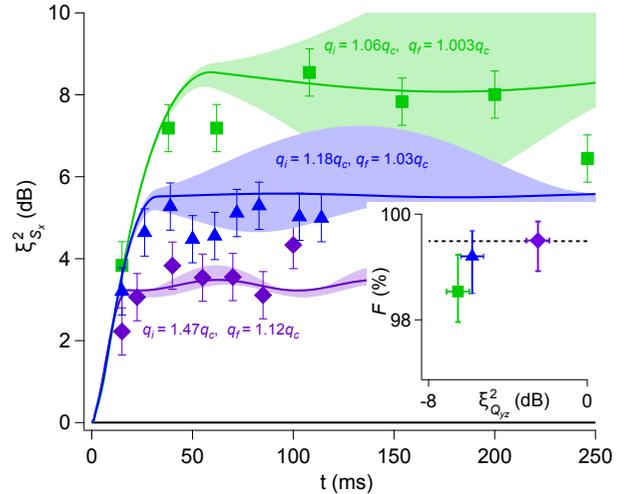}
\caption{\label{fig4} \textbf{Measurement of $\xi^2_{S_x}$ versus $t$ following the double quench sequence  for different $q_f$}. The solid lines are simulation results and the shaded regions reflect the  sensitivity of the simulations to the uncertainty in $c=-8.5\pm 0.1$ Hz. For the $q_f=1.003q_c$ data (green squares), the uncertainty of $c$ may lead to crossing over to the FM phase. Inset: the fidelity of the ground state $F$ determined from the residual oscillation of $\xi^2_{S_x}$ after the second quench. The maximum fidelity that can be detected (dashed line) is limited by the detection noise.
}
\end{figure}

The degree of squeezing  in the ground state increases as $q_f$ approaches $q_c$ according to Eq. \ref{squeezing_amount} because the semi-classical orbits near the pole become more elliptical  (Fig. 1). In Fig. \ref{fig4},   noise measurements are made for three different final $q_f$ values to show this dependency. We measure the anti-squeezed quadrature $\xi^2_{S_x}$ instead of the squeezing in $\Delta Q_{yz}$ to avoid limitations due to the detection noise limit. The sensitivity of the final state on the uncertainty in $c$ (and hence $q_c$)  increases at higher anti-squeezing amplitudes as shown by the shaded envelopes on the simulation curves.  Tomographic measurements shown in Supplementary Section II confirm that the maximum squeezing angle   $\theta_s=-\pi$ is  independent of $q_f$.

Following the second quench, any residual oscillation of the measured fluctuations $A=\left(\max(\xi^{2}_{S_x})-\min(\xi^{2}_{S_x})\right)/2$ is an indication of imperfect transfer into the ground state. Using a simple harmonic oscillator model \cite{xin2021fast}, and defining $F=\vert\langle \Psi(t)\vert\Omega\rangle\vert^2$ as the fidelity of the targeted ground state $\vert\Omega\rangle$ of $\hat{H}(q_f)$,  the fidelity  can be determined from the oscillation amplitude through: 

\begin{equation}
 F\approx 1- (A/2\eta)^2.
\label{correlation}
\end{equation}
Using this result, we determine that  $F>98\%$ for squeezed ground states as shown in Fig. \ref{fig4} inset. The tolerance to the oscillation is high because a small amount of excitation can lead to significant noise fluctuation. $F$ is lower at bigger $\xi^2_{S_x}$ because the sensitivity to $c$ robustness increases. The maximum fidelity that can be detected is limited by the noise detection uncertainty.
 
\begin{figure}
\includegraphics[width=0.5\textwidth]{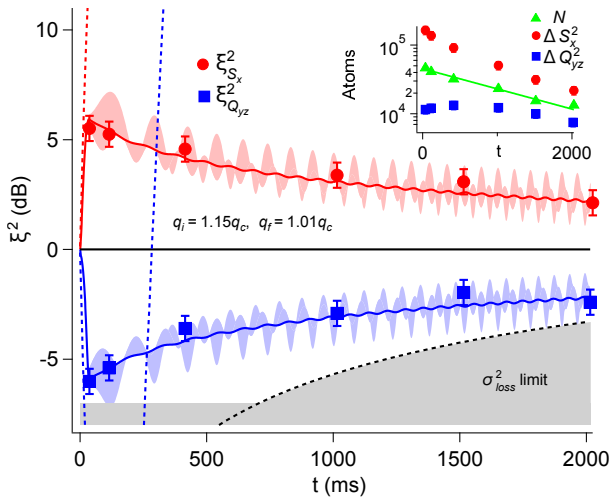}
\caption{\label{fig5} \textbf{Measurement of the long-term evolution of $\xi^2_{S_x}$ and $\xi^2_{Q_{yz}}$ in the squeezed ground state.} The simulations (solid lines) include the effects of atom loss $c(t)=(-8.7\pm 0.1)\exp(-2t/5\tau)$ Hz. Here $q_c=2\vert c(0)\vert $ is the critical point in the beginning of dynamics. The detection limit is dominated by the uncorrelated atom loss $\sigma^2_{loss}$ (black dashed line) after $600$ ms. The blue and red dashed lines are the maximum and minimum variance of the deep-quench squeezed state \cite{Hamley2012}.  The inset shows  $\Delta S^2_x$ (red circles), $\Delta Q^2_{yz}$ (blue squares) and $N$ (green triangles) versus $t$. 
}
\end{figure}

In Fig. \ref{fig5}, the long-term evolution of the squeezed ground state is measured. Atom loss due to the finite lifetime of the condensate leads to a decrease in peak density $n_0$, with $n_0 \propto N^{2/5}$ in the Thomas-Fermi model \cite{Gerving2012} (see Methods). This in turn affects the spinor dynamical rate and the QCP because  $q_c \propto c\propto n_0$. Hence, as the condensate decays, one expects that $q_f/q_c$ will increase, leading to a decrease in the squeezing. The data in the figure show this trend and compare well with simulations that include  exponential  atom loss with a time constant $\tau=3.2$~s thus leading to the attenuation of the squeezing amplitude. The ground state maintains squeezing for over $2$ s, and spin-noise tomography  shows that the minimum squeezing quadrature angle remains fixed at $\theta_{s,min}=-\pi$ throughout the entire evolution (see Supplementary Section II). The atom loss also degrades the squeezing due to uncorrelated atom loss \cite{PhysRevLett.107.210406}. This limit to the squeezing is also included in Fig. \ref{fig5} as ,$\sigma^2_{loss}$. The uncorrelated loss becomes more important at longer timescales comparable to the condensate lifetime. The inset shows directly the time evolution of the variances $\Delta S^2_x$ and $\Delta Q^2_{yz}$ together with the exponentially decaying total atom number, $N$.

The double quench method can be easily adapted to (pseudo) spin-1/2 systems such as bosonic Josephson junctions (BJJs) governed by a Hamiltonian of the form $\hat{H}=\alpha \hat{J_z}^2+J_x$ . It can also  be employed for spin-1 condensates with $c>0$ \cite{Zhao2014, Sala2016,PhysRevA.74.033612,doi:10.1126/science.abd8206,PhysRevLett.84.4031}  such as for sodium condensates. These systems have a QCP at $q=0$ but lack a continuous quantum phase transition. The result in this paper can be also extended to other systems similar to ours, such as bosonic Josephson junction systems \cite{Laudat2018} and the Lipkin-Meshkov-Glick model \cite{Solinas2008}.

In summary, this is the first realization of spin squeezed ground states in a spin-1 BEC within the proximity of the quantum phase transition point and provides a solid foundation for the application of our protocol. The result shows metrology improvements at a lifetime scale and the maintenance of the maximum squeezing angle in good agreement with theoretical predictions. Our method, implemented here near a second-order quantum phase transition, can also be used as a tool to measure the quantum phase transition precisely. This is exceptionally useful in condensed matter systems \cite{PhysRevB.44.11911}, for example, it can help answer the relationship between high-temperature superconductivity and the QCP in copper-oxide \cite{Broun2008,Gegenwart2008,Sachdev2008}. 

\section*{Online Content} Any methods, additional references, Nature Research reporting summaries, source data, extended data, supplementary information, acknowledgements, peer review information; details of
author contributions and competing interests; and statements of
data and code availability are available at (url)


\bibliography{sn-bibliography}


\clearpage
\section*{Methods}\label{sec: theoretical detail}
\bmhead{Initial state preparation} For the experiment, we prepare a condensate of $N=5 \times 10^4$ atoms in a cross optical dipole trap formed by a $\lambda=\SI{850}{\nano\meter}$ laser and a CO$_2$ laser ($\lambda=\SI{10.6}{\micro\meter}$) as illustrated in Fig. \ref{fig:subfig4}(a). The initial spin state is initialized in  $\vert F=1,m_F=0\rangle$  by applying a strong magnetic gradient during evaporative cooling, and the condensate is created in a  $B_0=1.1$ G  magnetic bias field. 

\bmhead{Calibration of collisional spin interaction} The collisional spin interaction energy $c$ is determined by careful measurement of the QCP using a quench technique  \cite{Anquez2016}. Quenching the condensate to fields close to the QCP and measuring the relative spin populations following $165$ ms of evolution at the final field, it is possible to determine $q_c$ with a precision of $\pm 0.1$ Hz (see Fig. \ref{fig:Bcrit}(a)). This same method is used to verify the relationship $c\propto N^{2/5}$ by studying $q_c$ as a function of atom number (see Fig. \ref{fig:Bcrit}(b)). For the experiments, $c$ ranged from $[-7.5,-8.7]$ Hz due to day-to-day variations of the experimental conditions.

\bmhead{Phase encoding} Measurements of $\Delta S^2_x$ are performed by first doing a radio-frequency (RF) pulse of $\exp(-i\hat{S}_x\pi/2)$ in the spin-1 manifold (Fig. \ref{fig:subfig4}(b)). 
The $Q_{yz}$ direction measurement is done with a microwave (µwave) pulse $\exp(-i\hat{Q}_{zz}\pi/4)$ detuned from the clock transition to first shift the spinor phase by $\Delta\theta_s=-\pi$ followed by the RF rotation \cite{Hamley2012, PhysRevLett.111.090403}. To be able to shift the spinor phase $\theta_s$ precisely, the quadratic Zeeman effect needs to be accounted for. The hyperfine splitting is calculated using the Breit-Rabi formula \cite{steck2001rubidium}. The clock transition between $\vert F=1,m_F=0\rangle$ and $\vert F=2,m_F=0\rangle$ has the energy difference
\begin{equation}
\begin{split}
\Delta_{E}&=E_{20}-E_{10}\approx E_{hfs}+\frac{1}{2}\frac{(g_J \mu_B)^2}{E_{hfs}}B^2\\
&=6834682610.9\text{ Hz}+572.8\text{ Hz/G}^2\cdot B^2.
\end{split}
\end{equation}
The resonance of the clock transition needs to be adjusted depending on the magnetic field $B$.

\bmhead{SQL measurement} The atom detection is calibrated using a coherent RF rotation to measure the standard quantum limit (SQL) \cite{PhysRevLett.107.210406}. The calibration is performed at the same final magnetic field as the squeezing measurements $4$ ms after a fast quench to minimize spin evolution \cite{Hamley2012}.    The quantum projection noise $\sigma^2_{QPN}=\Delta M^2-\sigma^2_{PSN}-\sigma^2_{bkg}$ is extracted by subtracting the photon shot noise $\sigma^2_{PSN}$ and the background imaging noise $\sigma^2_{bkg}$ from the measured magnetization variance $\Delta M^2$ \cite{PhysRevLett.107.210406}. The uncertainty of $\sigma^2_{QPN}$ is given by $\mbox{std}(\sigma^2_{QPN})=\sigma^2_{QPN}\sqrt{\frac{2}{N_s-1}}$, where  $N_s$ is the number of measurements.

\bmhead{Magnetic field gradient cancellation} Empirically, we have found that in order to observe well-characterized spin dynamical evolution, it is necessary to zero the magnetic field gradient along the $z$ axis (CO$_2$ laser axis), as shown in Fig. \ref{supfig3}(c). To measure the gradient, we perform magnetic field measurements using the condensate careful translated to different $z$ position. We use a motorized translation stage to precisely control the spatial location of the condensate by changing the CO$_2$ laser trap focus point. The range of translation is measured via absorption imaging to be $150$ µm. At each location, RF spectroscopy is performed to measure the local magnetic field (see Fig. \ref{supfig3}(a) for a typical measurement). A linear fit to the data is used to determine the magnetic gradient.   We use auxiliary anti-Helmholtz coils near the chamber to cancel this gradient.  Fig. \ref{supfig3}(b) shows that we can cancel the gradient to  $<10$ mG/cm.

The  squeezing detection limit of $-7$ dB is determined from measurement of spin-mixing number squeezing based on \cite{PhysRevLett.107.210406}. Spin-mixing (Fig. \ref{supfig3}(c)) generates correlated pairs of atoms in $m_F=\pm 1$ that exhibit number squeezing in the magnetization $ M = N_{+1} - N_{-1}$, similar to  optical four-wave mixing. The magnetization variance is measured by counting the spin populations following Stern-Gerlach separation to determine the maximum detectable squeezing.

\section*{Data availability}
The data that support the findings of this study are available from the corresponding author upon reasonable request.

\section*{Code availability}
The codes used for simulation and analysis are available from the corresponding
author upon reasonable request.

\section*{Acknowledgments}
We would like to thank T. A. B Kennedy and C.A.R. Sá de Melo for fruitful discussions and insights. We also acknowledge support from the National Science Foundation, grant no. NSF PHYS-1806315.

\section*{Author contribution}
L. X. and M. S. C. conceived this study. L. X., M. B. and J. T. C. performed the
experiment and analysed the data. L. X. conducted the numerical
simulations. L. X., M. B., J. T. C. and M. S. C wrote the paper.

\section*{Competing interests}
The authors declare no competing interests.

\section*{Additional information}
\textbf{Supplementary information} The online version contains supplementary material
available at 

\noindent\textbf{Peer review information}

\noindent\textbf{Correspondence and requests for materials} should be addressed to Michael S. Chapman.

\noindent\textbf{Reprints and permissions information} is available at www.nature.com/reprints.

\renewcommand\thefigure{M\arabic{figure}} 
\setcounter{figure}{0} 
\begin{figure*}
\begin{center}
\includegraphics[width=0.5\textwidth]{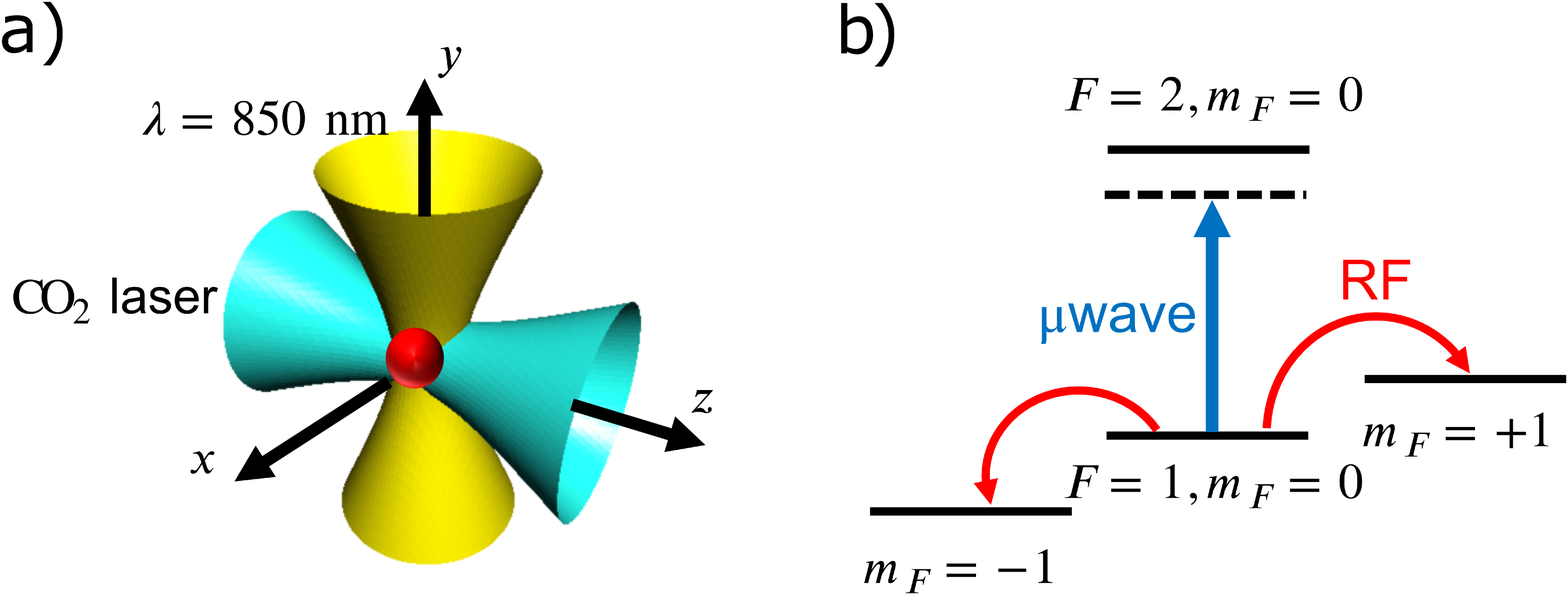}
\caption{\label{fig:subfig4}(a) An illustration of our apparatus.  The optical trap is formed by a CO$_2$ laser ($\lambda=\SI{10.6}{\micro\meter}$) in horizontal direction and a $\lambda=\SI{850}{\nano\meter}$ laser in vertical direction. (b) The spin state tomography is performed using the RF pulses and the detuned µwave pulses as shown in the figure.
}
\end{center}
\end{figure*}

\begin{figure*}
\begin{center}
\includegraphics[width=0.5\textwidth]{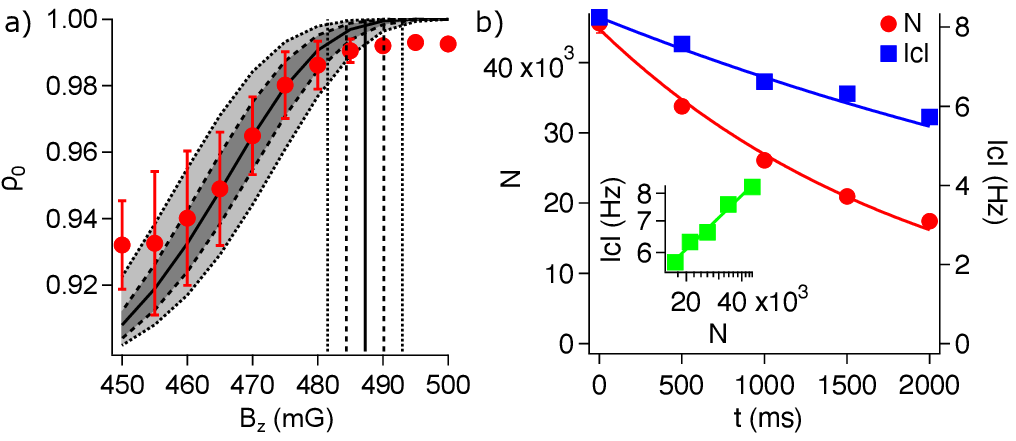}
\caption{\label{fig:Bcrit}(a) The critical point detection using sudden quench $B_z$ scans at $t=165$ ms. $\rho_0$ is measured as a function of $B_z$. 1\% pollution from $\lambda=850$ nm laser is presented compared to the simulation result as discussed in Fig. \ref{fig3}. The black solid line $c=-8.5$ Hz, dashed lines $\pm0.1$ Hz, and dotted lines $\pm0.2$ Hz are attached to the figure. By using the simulation, $c$ can be decided with precision about $\pm0.1$ Hz. (b) Different atom number $N$ generated by a finite life-time decay of BECs versus $c$ is studied using the method in (a). The measured $|c|$ agrees well with the analytic relationship $c\propto N^{2/5}$ (inset green line). }
\end{center}
\end{figure*}

\begin{figure*}
\begin{center}
\includegraphics[width=0.45\textwidth]{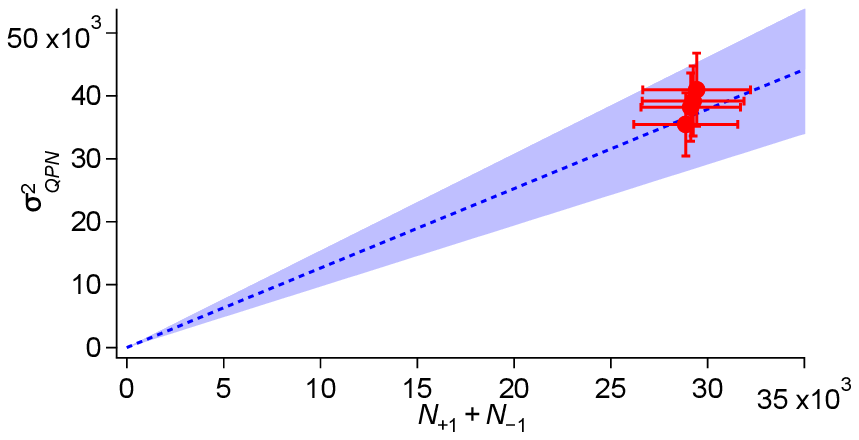}
\caption{\label{subfig1} The atom detection is calibrated to measure the SQL. The calibration is performed $t=4$ ms after a fast quench to minimize spin evolution at the same final magnetic field. Theoretical prediction gives 0.87 dB($\pm1$ dB) anti-squeezing. The count per atom with 200 µs exposure time is 157.9 counts/atom. 
}
\end{center}
\end{figure*}

\begin{figure*}
\begin{center}
\includegraphics[width=0.5\textwidth]{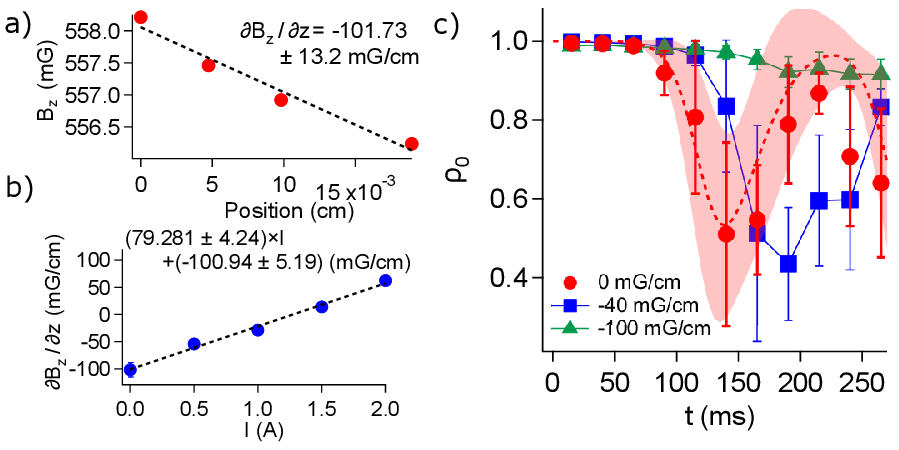}
\caption{\label{supfig3}(a) Local $B_z$ is measured for the condensate carefully translated to different $z$ position, and $\partial B_z/\partial z$ is determined through a linear fit to the data. Estimated error of the magnetic field from RF spectroscopy is smaller than the size of markers. (b) Measurement of $\partial B_z/\partial z$ versus the current $I$ in auxiliary anti-Helmholtz coils. We use those coils to cancel the ambient gradient to $<10$ mG/cm.  (c) Spin dynamical evolution (spin-mixing) at $B_z=250$ mG, 10 shots for each point. Correlated pairs of $m_F=\pm1$ are generated, similar to the optical four-wave mixing. The data agrees with the mean-field simulation prediction only with a precisely canceled $\partial B_z/\partial z$ (red circles). 
}
\end{center}
\end{figure*}




\end{document}


\preprint{APS/123-QED}

\title{Supplementary Information: Squeezed Ground States in a Spin-1 Bose-Einstein Condensate}

\author{Lin Xin}\author{Maryrose Barrios}\author{Julia T. Cohen}\author{Michael S. Chapman}
\affiliation{%
School of Physics, Georgia Institute of Technology, Atlanta, GA 30332, U.S.A\\
}%

\maketitle

\section{Theoretical details}\label{sec: theoretical detail}
 \begin{equation}
\hat{H}=\frac{c}{2N}\hat{S}^2-\frac{q}{2}\hat{Q}_{z},
\label{hamiltonianreduced}
\end{equation}
The Hamiltonian in Eq. \ref{hamiltonianreduced} is written in terms of many-body operators, defined as follows. The operator in the two-body s-wave collision term is the square of the collective spin operator, $\hat{S}^2=\hat{S}^2_{x}+\hat{S}^2_{y}+\hat{S}^2_{z}$. The $\nu$ component is $\hat{S}_\nu=\sum^N_{i=1} \hat{s}^i_{\nu}$, and $\hat{s}^i_\nu$ is the corresponding single body spin-1 operator for the $i^{\text{th}}$ particle. The operator $\hat{Q}_{z}\equiv-\hat{N}/3-\hat{Q}_{zz}$ is simply related to a collective nematic (quadrupole)  tensor $\hat{Q}_{\nu\mu}=\sum^N_{i=1} \hat{q}^i_{\nu\mu}$, where $\hat{q}_{\nu\mu}\equiv\hat{s}_{\nu}\hat{s}_{\mu}+\hat{s}_{\mu}\hat{s}_{\nu}-(4/3)\delta_{\nu\mu}$ is the single particle symmetric and traceless tensor. $q=q_z B^2$ is the quadratic Zeeman energy per particle in an applied magnetic field $B$ with $q_z / h \approx 71.6$~Hz/G$^2$ (hereafter, we set Planck's constant $h=1$). The quantum spin states and their evolution within the $\hat{S}_z=0$ subspace can be visualized with the  unit spheres of the $\{S_{x}, Q_{yz}, Q_z\}$ and $\{S_{y}, Q_{xz}, Q_z\}$ variables. Since the two unit spheres share the same dynamics up to the irrelevant Larmor phase, we focus only on the $\{S_{x}, Q_{yz}, Q_z\}$ sphere for the figures. The QCP between the polar and the FM phases occurs at $q_c=2|c|$. This is a second-order (continuous) quantum phase transition according to Ehrenfest’s classification, which is akin to the phase transition in the Landau-Ginzburg model \cite{ZUREK1996177}. The corresponding polar phase ($q>q_c$) energy gap $\Delta(q)=\omega(q)=\sqrt{q(q-q_c)}$ between the ground state and the first excited state vanishes at the QCP in $N\to+\infty$ case.

\begin{figure}
\includegraphics[width=0.45\textwidth]{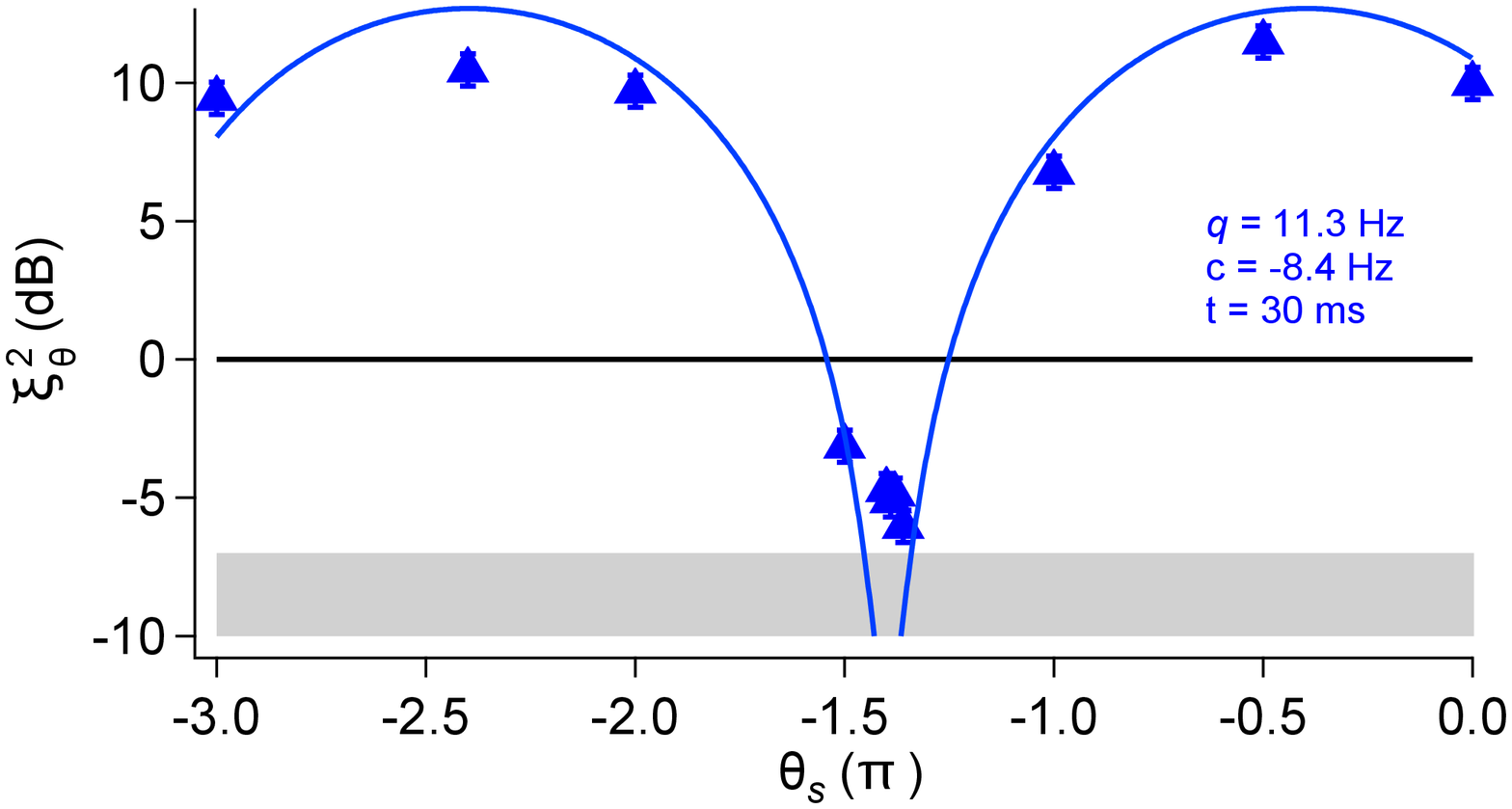}
\caption{\label{fig:subfig2b} Spin-noise tomography measurement for the deep quench method at $B_z=400$ mG, $t=30$ ms. The simulations (blue solid line) match well without any adjustments. The maximum squeezing does not have a fixed  $\theta_{s,min}$ because it is an asymptotic function of parameters $q/|c|$ and $t$.
}
\end{figure}

We numerically solve the full quantum spin-1 dynamics in the single-mode approximation
\cite{Law1998},
$$
i\hbar\partial_t|\Psi(t)\rangle=\hat{H}|\Psi(t)\rangle$$  using the $\hat{S}_z = 0$ Fock state basis $|N_{1},N_{0},N_{-1}\rangle=|k,N-2k,k\rangle =:|k\rangle$, $0 \leq k \leq \frac{N}{2}$. In this basis, the non-zero  matrix elements of the Hamiltonian matrix are
\begin{align*}
\langle k'|\hat{S}^2|k \rangle &=2\bigg[\Big(2(N-2k)k+(N-k)\Big)\delta_{k',k}+\\
&(k+1)\sqrt{N-2k}\sqrt{N-2k-1}\delta_{k',k+1}+\\
&(k)\sqrt{N-2k+2}\sqrt{N-2k+1}\delta_{k',k-1}\bigg],\\
\langle k'|\hat{Q}_{z}|k \rangle &=4k \delta_{k',k},
\end{align*}
and the initial condition is $|\Psi(t=0)\rangle=|0,N,0\rangle$. In the $N=+\infty$ limit, the dynamics can be described by the mean-field equations \cite{PhysRevA.72.013602}:
\begin{align*}
\dot{\rho}_0&=\frac{2c}{\hbar}\rho_0\sqrt{(1-\rho_0)^2-m^2}\sin(\theta_s)\\
\dot{\theta_s}&=-\frac{2q}{\hbar}+\frac{2c}{\hbar}(1-2\rho_0)\\
&\ \ \ \ \ \ \ +\frac{2c}{\hbar}\frac{(1-\rho_0)(1-2\rho_0)-m^2}{\sqrt{(1-\rho_0)^2-m^2}}\cos(\theta_s),
\end{align*}
where $\rho_0=N_0/N$ is the relative population of $|m_F=0\rangle$, $m=(N_{+1}-N_{-1})/N_0$ is the relative magnetization, and $\theta_s=\theta_{+1}+\theta_{-1}-2\theta_0$ is the relative spinor phase defined in terms of the phases of the Zeeman components. The initial ensemble is defined to satisfy the quantum uncertainty relationships $\Delta S_x\Delta Q_{yz}=N$ and $\Delta S_y\Delta Q_{xz}=N$. The simulations agree with the experimental data qualitatively but require a correction of $\xi^2 = (\xi^2_{sim})^{0.7}$ to match quantitatively. 
Interestingly, this correction is only required to match the squeezing measurements in the neighborhood of the QCP. For squeezed state generation using the deep quench method, the simulations match well without any adjustments, as shown in Fig. \ref{fig:subfig2b}. We do not currently understand this discrepancy; perhaps normally negligible effects such as  magnetic anisotropy \cite{PhysRevLett.84.4031} or dipolar interactions \cite{PhysRevLett.100.170403} become significant  near the critical point where the energy scale goes to zero. We hope to further investigate this in the future.

\begin{figure*}
\includegraphics[width=\textwidth]{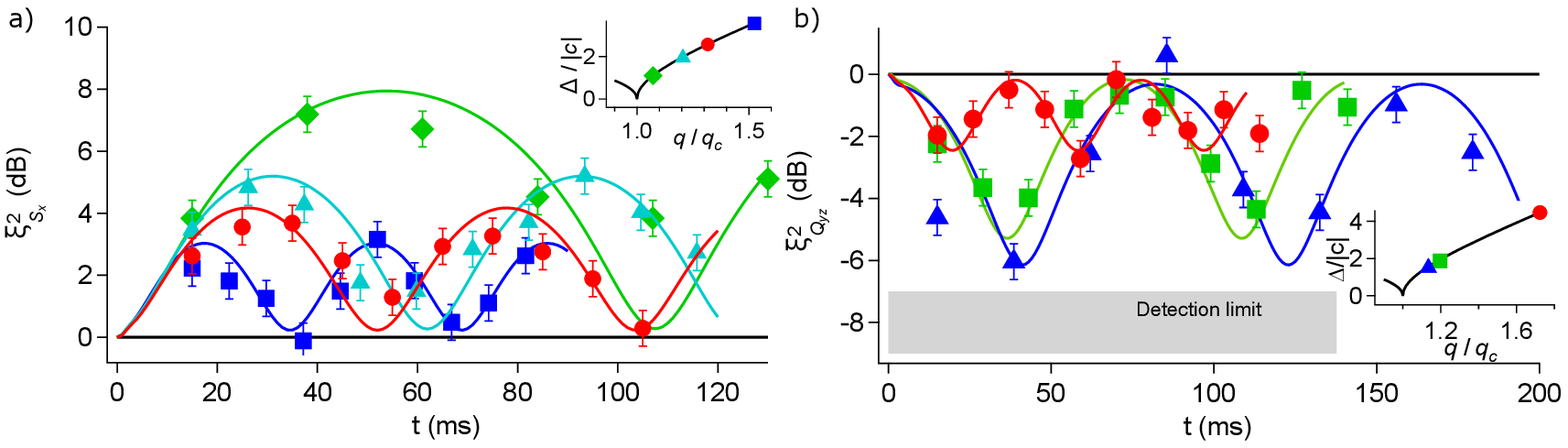}
\caption{\label{supfig5}The periodic quadrature variance data of (a) $S_x$ and (b) $Q_{yz}$. The insets show the relative distance from the QCP. As $q_i$ approaches $q_c$, $|\xi^2_{S_x}|$ and $|\xi^2_{Q_{yz}}|$ become bigger while $\omega_i$ gets smaller. Solid curves are the simulation result with corrections. Inset: The measured oscillation frequency from the data is converted into the energy gap $\Delta=\hbar\omega_i$ and it measures the distance between $q_i$ and $q_c$. }
\end{figure*}

\section{Supplementary measurements}\label{section:nonequ}
In the main context, we studied the periodic squeezing for $q_i=1.16q_c$. We have also studied the periodic squeezing for different $q_i$, as shown in Fig. \ref{supfig5}. The temporal evolution rotates the distribution and creates periodic squeezing described by
\begin{equation}\label{oscilamp}
    \xi_{Q_{yz}}^2=\frac{1+\cos(\omega_i t)}{2}+\frac{1-\cos(\omega_i t)}{2\eta}
\end{equation}
and $\xi_{S_x}^2=1/\xi_{Q_{yz}}^2$. As $q_i$ approaches $q_c$, $|\xi_{S_x}^2|$ and $|\xi_{Q_{yz}}^2|$ become bigger while $\omega_i$ gets smaller.

\begin{figure}
\includegraphics[width=0.45\textwidth]{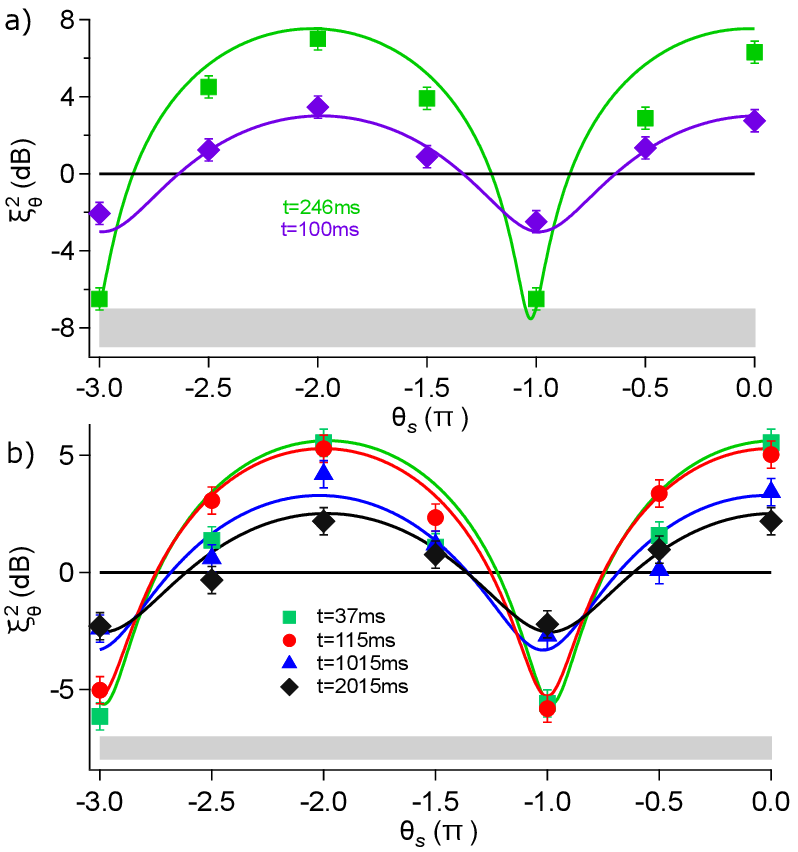}
\caption{\label{fig:tomography}(a) Spin-noise tomography for stabilized squeezing with $q_f=19.1$ Hz, $t=100$ ms and $q_f=17.05$ Hz, $t=246$ ms. (b) Spin-noise tomography for states held inside the trap for $37,115,1015,2015$ ms shows that $\theta_{s,min}=-\pi$ is fixed during the entire evolution.}
\end{figure}

The extra spin-noise tomography data for the stabilized squeezing in the main context is plotted in Fig. \ref{fig:tomography}. It is evident that the squeezed ground state has a fixed $\theta_{s,min}=-\pi$ that is independent of $q_f$ and $t$. One of the attractive features of our method is that no searching is needed to align the minimum squeezing direction to the detection variables.

\begin{figure}
\includegraphics[width=0.45\textwidth]{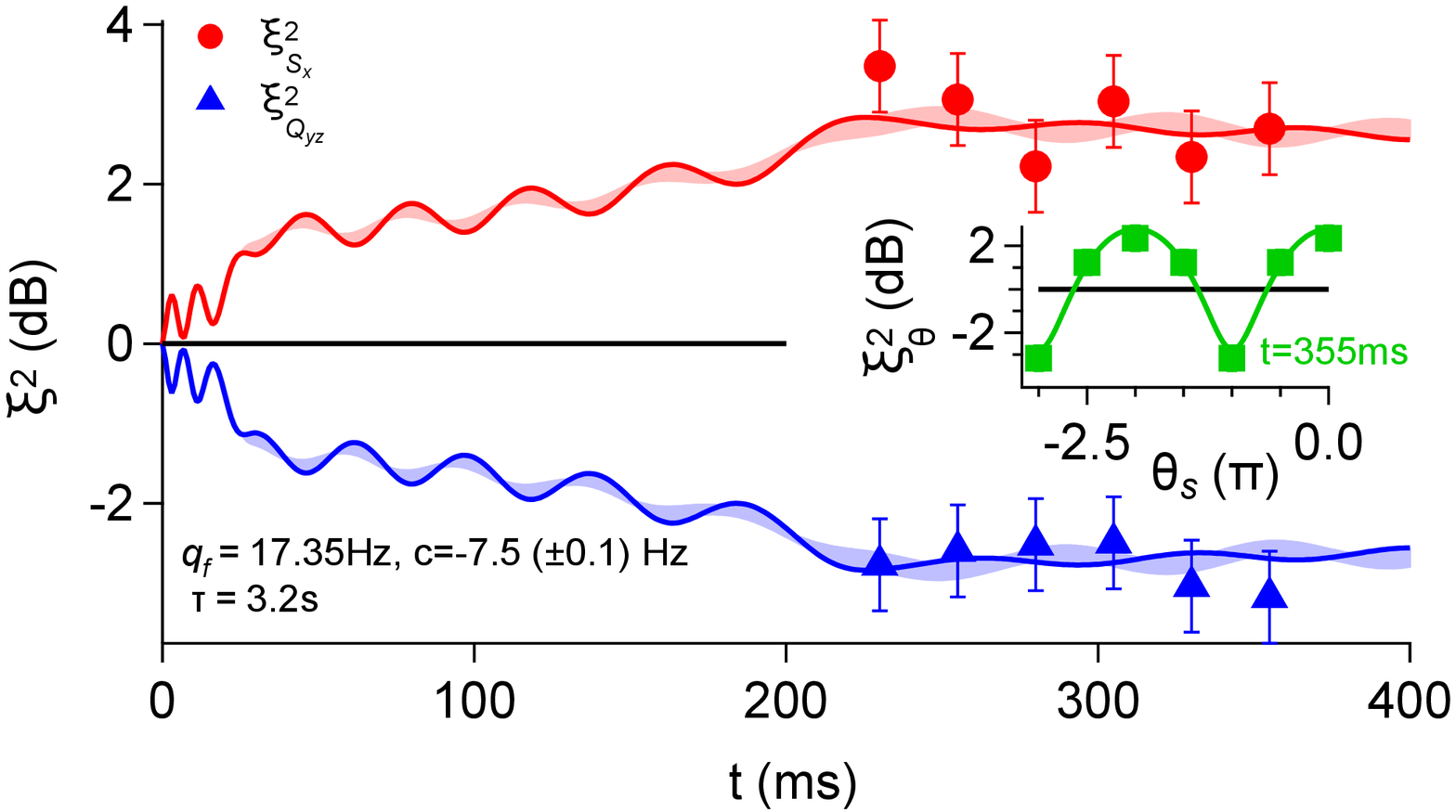}
\caption{\label{fig6}A squeezed ground state is generated with an adiabatic ramp. The adiabatic passage is more robust but slower compared to the shortcut protocol. The inset shows the spin-noise tomography measured at $t=355$ ms. The simulations take the lifetime of the BEC into consideration with $\tau=3.2$ s. 
}
\end{figure}
\begin{figure}
\includegraphics[width=0.45\textwidth]{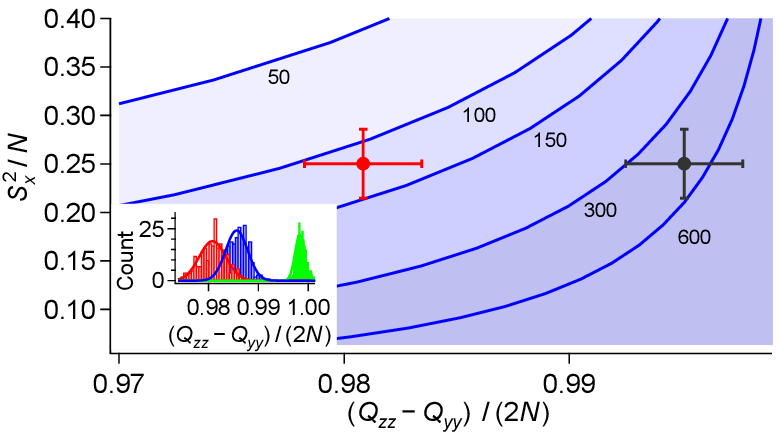}
\caption{\label{fig3}A non-separable subset of 150 particles (red dot) is detected in the squeezed ground state based on the entanglement breadth measurement. The corrected spin length based on $\lambda=850$ nm laser pollution (black dot) shows a 600 particles entanglement. Inset: Measurement of the spin length under different experimental conditions. Green bars are the measurement in the CO$_2$ trap which sets the detection limit for the maximum spin length. Blue bars are the data inside the cross trap at the condition $q\gg q_c$. The red bars are the measurements for a $-6$ dB squeezed ground state. 
}
\end{figure}

A high fidelity $-3$ dB squeezing generated by the adiabatic passage is measured in Fig. \ref{fig6}. The assumption of constant atom number is no longer valid at timescale $>100$ ms due to the finite lifetime ($\tau=3.2$ s) of our condensates. The simulation with $c(t)=-7.5\exp\big(-2t/(5\tau)\big)$ Hz is plotted in Fig. \ref{fig6}. As the time increases, the spin correlations loss $\sigma^2_{loss}=p(1-p)N^0$ \cite{PhysRevLett.107.210406} due to the lifetime leads to smaller detecable squeezing, where $N^0$ is the number of atoms without loss and $p=1-\exp(-t/\tau)$ is the probability of atom loss.  A two-step linear ramp is used to realize the adiabatic passage. The adiabatic passage ramp takes $30$ ms to ramp from $B_0$ to $600$ mG and then takes $200$ ms to ramp to $B_f$.

We measure the entanglement of squeezed ground states through the entanglement breadth as shown in Fig. \ref{fig3}. The entanglement breadth in the basis of the spin-nematic operators can be calculated in analogy with the Bloch sphere operators \cite{PhysRevLett.112.155304,Zou6381}. The boundary labeled by the number $k$ is given by the state 
\begin{equation}
    |\Psi\rangle=|\psi_k\rangle^{\otimes n}\otimes |\psi_p\rangle
\end{equation}
which is a product of $n$ ($=[N/k]$, integer part of $N/k$) copies of state $|\psi_k\rangle$ containing $k$ nonseparable spin-1 particles and state $|\psi_p\rangle$ composed of the remaining $p$ $(=N-nk)$ particles. The state $|\psi_\mu\rangle\ (\mu=k,p)$ represents the ground state of the $\mu$ particles Hamiltonian
\begin{equation}
    H_\mu=(\hat{S}_x^{(\mu)})^2-\lambda (\hat{Q}^{(\mu)}_{zz}-\hat{Q}^{(\mu)}_{yy})/2
\end{equation}
The boundary points are obtained as 
\begin{equation}
\begin{split}
    \langle (\hat{Q}_{zz}-\hat{Q}_{yy})/2 \rangle=n&\langle (\hat{Q}^{(k)}_{zz}-\hat{Q}^{(k)}_{yy})/2 \rangle_{|\psi_k\rangle}\\
    &+\langle(\hat{Q}^{(p)}_{zz}-\hat{Q}^{(p)}_{yy})/2 \rangle_{|\psi_p\rangle}
    \end{split}
\end{equation}
\begin{equation}
    \begin{split}
        (\Delta \hat{S}_x)^2&=n(\Delta \hat{S}^{(k)}_x)^2_{|\psi_k\rangle}+(\Delta \hat{S}^{(p)}_x)^2_{|\psi_p\rangle} 
    \end{split}
\end{equation}
The spin length here is different from the Dicke state's case because the spin vector is well-pointed in the $(Q_{zz}-Q_{yy})/2$ direction.

In the inset of Fig. \ref{fig3}, the spin length $|(Q_{zz}-Q_{yy})/2|=2N_0-N$ is measured and studied for different experimental conditions. The $\lambda=850$ nm laser contamination at the condition $q\gg q_c$ leads to a $1\%$ fraction of atoms in $m_F=\pm1$ compared to the ideal case in the sole CO$_2$ laser trap. The squeezed ground state at $q_f\gtrsim q_c$ further reduces the spin length by $1\%$. A non-separable subset of 150 (600 with correction) particles is detected in the squeezed ground state in Fig. \ref{fig3}.

\newpage
\bibliography{apssamp}